**Reply to comment "Divergent and Ultrahigh Thermal Conductivity in Millimeter-Long Nanotubes":**

We regret that PRL did not accept our Reply for publication. We believe that both our Reply and the original paper is sound and correct. We post our reply here and let readers to judge.

The comment by Li *et al*. [1] has two points: (1) the temperature profile of the heater is not parabolic, and (2) that radiation heat loss from the single-wall carbon nanotube (SWCNT) induces an overestimation of the thermal conductivity of the sample. Here we show that Li *et al*. have confused two different measurement methods and misidentified our method to be similar to theirs. Therefore (1) introducing the non-parabolic correction by Li *et al*. has negligible (<0.1%) effects to our results; and (2) our measurements in fact *underestimate* the thermal conductivity of the SWCNTs, as emphasized in our paper [2].

There are two different experimental methods for measuring thermal conductivity of a sample. First, one can supply a constant power ($P_h$) to a heater and then measure its temperature rise before ($\Delta T_{h,before}$) and after ($\Delta T_{h,after}$) connecting it to a sample. The measured thermal conductance ($K_m$) of the sample is obtained using

$$K_m = P_h \left( \frac{1}{\Delta T_{h,after}} - \frac{1}{\Delta T_{h,before}} \right) \qquad (1)$$

As shown in Fig. 1(a), because only one probe is used for simultaneous heating and sensing, we dub it "one-probe method". Many scanning thermal microscopes [3,4], Prof. X. Zhang and Prof. K. Takahashi's previous works [5-8], optical techniques [9-11], and one of our previous works (see Methods I & II in Ref. [12]) have employed this method. Similar to Ref. [11], in which $\Delta T$ was measured using Raman shifts and $P_h$ was obtained from the laser absorption coefficient of a SWCNT, experiments using one-probe method commonly employ a "source $P_h$, measure $\Delta T$" measurement scheme.

On the other hand, many other experiments had incorporated an independent heater and an independent sensor, as displayed in Fig. 1(b), for measuring nanowires [13-15], nanotubes [16-18], or graphene [19,20]. Here $K_m$ is obtained using:

$$K_m = \frac{1}{\Delta T_h - \Delta T_s} \left( \frac{P_h \Delta T_s}{\Delta T_h + \Delta T_s} \right) = \frac{P_s}{\Delta T_h - \Delta T_s} \qquad (2)$$

where $\Delta T_h$ and $\Delta T_s$ is temperature rise of the heater and the sensor, respectively. Note that the term in the bracket denotes the fraction of the total heater power received by the sensor. In our work, $\Delta T_h - \Delta T_s$ was kept constant and thermal current flowing through the sensor was measured (i.e. $P_s = 2K_s\Delta T_s$, where $K_s$ is the thermal conductance of the sensor beam). This method is dubbed "two-probe method", using a "source $\Delta T$, measure $P_s$" scheme. Unlike the one-probe method in which the heater must be located at the ends of a multiprobe device, the two-probe method has no such limitation.

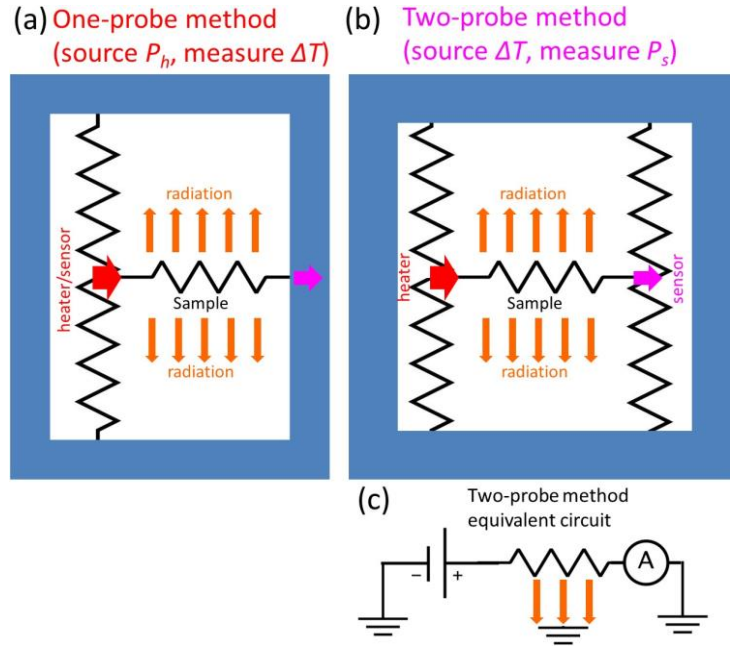

Fig. 1 (a, b) Two different experimental methods for measuring thermal conductance of a sample. (c) An equivalent electrical circuit diagram of (b).

Now we discuss how would the radiation heat loss from a SWCNT make $K_m$ deviate from the intrinsic value of thermal conductance ($K$). Note that in Figs. 1(a & b), we always have $P_h > P_s$ whenever there is radiation heat loss from the sample. However, because the "source $P_h$, measure $\Delta T$" scheme is used in Fig. 1(a), it results in $K_m > K$. On the other hand, we had employed a "source $\Delta T$, measure $P_s$" scheme in Fig. 1(b), thus we concluded $K_m < K$ [2]. In fact, our method is equivalent to a two-probe electrical resistance measurement using a "source $V$, measure $I$" scheme, as shown in Fig. 1(c). Readers can easily verify our statements by analyzing the circuit.

Likewise, because the measured thermal conductance of our 1mm-long SWCNT is $1.77 \pm 0.15 \times 10^{-11}$ W/K and the thermal conductance of the heater beam is more than $2 \times 10^{-8}$ W/K, the non-parabolic correction to Eq. (2) is smaller than 0.1% and will not affect our conclusions.

In the Supplemental Material, the definition of $R_{bi}$ (page 3, after Eq. S4) should be corrected to "$R_{bi} = 2L/\kappa_{bi}A$ is the total thermal resistance of the $RT_i$, *measured from one end to the other end of the beam*". The correction does not make any changes to our calculations or results.

Lastly, regarding to Liu *et al.*'s experiment that suggests a convergent thermal conductivity of SWCNTs at length (*L*) ~10μm [11], we must point out that the absence of error bars in their data has much perplexed us. Because the same technique demonstrated by other groups have shown to exhibit >10% uncertainty in the temperature measurement [10,21,22], and, additionally, complex position dependent variations [10,22], these would render Liu *et al.*'s data inconclusive for $L > 5$μm.

We thank Cheng-Li Chiu's helps on preparing the reply.

Victor Lee,[1,2] Chi-Hsun Wu,[1,2] Zong-Xing Lou,[1,2] Wei-Li Lee,[3] and Chih-Wei Chang[1]

[1] Center for Condensed Matter Sciences, National Taiwan University, Taipei 10617, Taiwan
[2] Department of Physics, National Taiwan University, Taipei 10617, Taiwan
[3] Institute of Physics, Academia Sinica, Taipei 11529, Taiwan